\documentclass{moriond}

\def\be{\begin{equation}}
\def\ee{\end{equation}}
\def\bea{\begin{eqnarray}}
\def\eea{\end{eqnarray}}

\newcommand{\Photo}{\includegraphics[height=35mm]{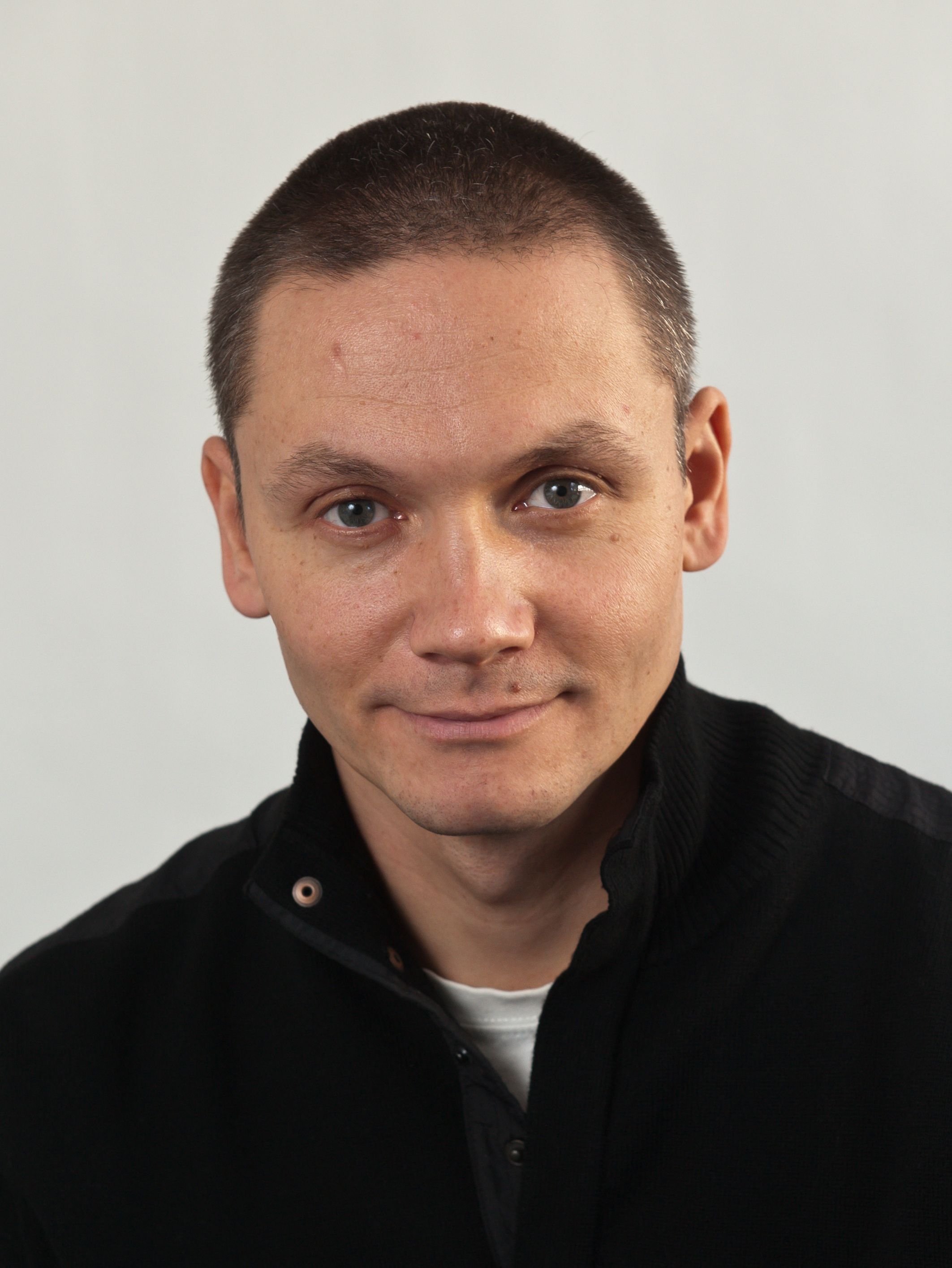}}

\begin{document}
\vspace*{4cm}
\title{MORIOND QCD AND HIGH ENERGY INTERACTIONS 2017: THEORETICAL SUMMARY}

\author{ OLEG LEBEDEV}

\address{Department of Physics, University of Helsinki, Gustaf H\"allstr\"omin katu 2a,
Helsinki, Finland }

\maketitle\abstracts{
I summarize the highlights of the theory talks at  the 52nd Rencontres de Moriond: QCD and high energy interactions.  }

\section{Preface}

In the LHC era, progress in particle physics is contingent on understanding the
LHC data,  for  which QCD is the crucial ingredient.   The role of QCD is twofold:
it is a fundamental theory important in its own right, which also allows us 
to  extract   fundamental parameters of Nature ($\alpha_s, m_t$,...)   from complicated data.
On the other hand, it is an inevitable ``background''  in searches for New Physics.
While in the earlier days of Moriond, the focus was placed on the first aspect, nowadays
both roles are equally important. (Presumably, this trend is reflected in the conference name change:
``QCD  and hadronic interactions'' $\rightarrow$  ``QCD  and
high energy interactions'' which took place in 2008.)  
At Moriond QCD 2017,  the two   aspects of QCD were  
given comparable significance.
The theory talks could be divided into  four broad categories: Higgs physics,
precision calculations, flavor physics and new phenomena (a.k.a. beyond the Standard Model).
Even though
such a division is quite ambiguous,  I find it convenient and will follow it in my summary. For further details, I encourage the reader to consult the original talks at  http://moriond.in2p3.fr/QCD/2017/MorQCD17Prog.html and references therein.

\section{Higgs Physics}

Physics of the Higgs sector in various incarnations  has   been one of the most popular subjects of the theory talks at  this conference. Although the consistency between the Standard Model (SM) predictions and the data is excellent (Fig.1), detectable imprints of new physics on the Higgs properties are well motivated. In particular, the Higgs field (unlike other Standard Model fields) can couple directly to the hidden sector  at the renormalizable level which would affect the Higgs couplings. The hidden sector can hold the key to the problems of dark matter and inflation which makes the Higgs a unique cosmological probe. 
 Pinning down the resulting  deviations in the Higgs couplings  requires good control over the SM predictions and thus precision calculations.

The theory talks have covered the following topics:

\begin{itemize}
\item  Higgs self--coupling 
\item  Higgs effective (EFT) couplings
\item  beyond the SM Higgses
\item  CP properties in multi--Higgs models
\item dark matter in multi--Higgs models
\item Higgs as the only fundamental scalar in Nature
\end{itemize}

Measuring the Higgs self--coupling is the next big challenge at the LHC. Even if the other Higgs couplings are close to their SM predictions, the self--coupling can deviate by ${\cal O}(100\%)$.
This could, for instance,  be due to the Higgs mixing with another (SM singlet) scalar or radiative effects of the hidden sector. The most straightforward channel sensitive to the Higgs triple coupling is di--Higgs production in gluon fusion, which proceeds through the $hhh$--vertex. The background for this process includes a box diagram with the top quark in the loop. Jones reported  on incorporating the full $m_t$-dependence of such diagrams  at NLO QCD, which improves significantly over the Higgs EFT ($m_t \rightarrow \infty$) limit
and leads to about $14\%$ correction to Higgs pair production.

\begin{figure}[t]
\begin{center}
\includegraphics[height=90mm]{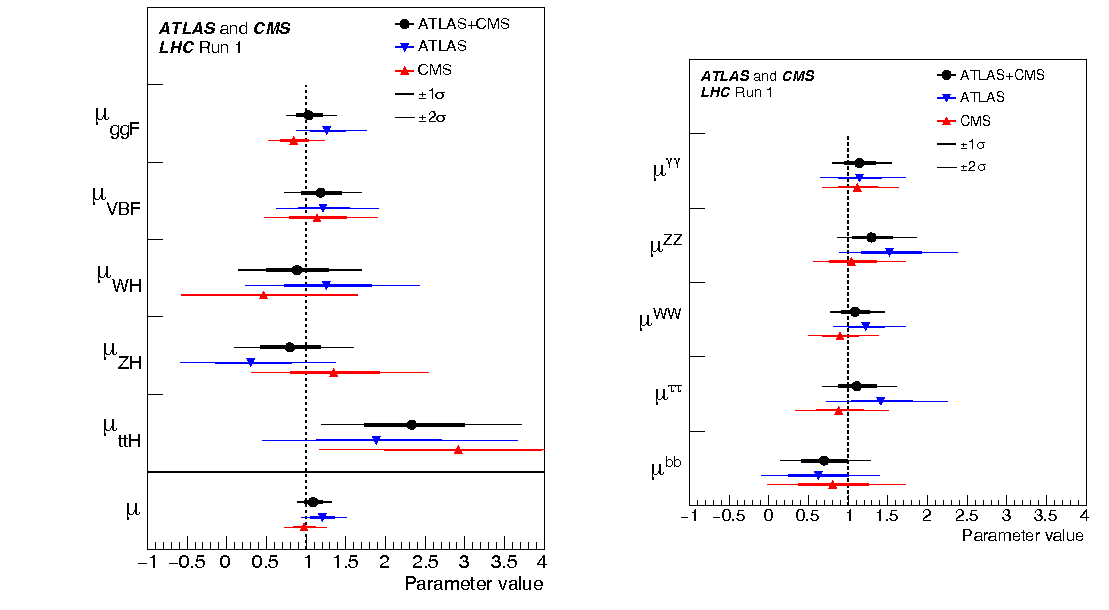}
\caption{Higgs production and decay signal strengths in different channels versus the Standard Model prediction ($\mu_i=1$). The figure is taken from  arXiv:1606.02266.}
\end{center}
\end{figure}

New physics in the Higgs sector may also manifest itself through dimension-6 effective operators of the type
\begin{equation}
{\cal O}_1 = \vert H \vert^2 G_{\mu\nu}^a G^{\mu\nu,a} ~,~ {\cal O}_2 = \vert H \vert^2 \bar Q_L H^c u_R + {\rm h.c.}~,~...
\end{equation}
These are obtained by integrating out heavy new physics states. 
Such operators   generally affect the Higgs production cross section. When multiple operators are present simultaneously, they can conspire to give rise to the same production rate as that in the SM.
Ilnicka discussed their   effects on the Higgs $p_T$ distribution which provides an additional handle on the size of these operators and breaks the above degeneracy.

Extensions of the Higgs sector were reviewed by Carena, with the focus on 2 Higgs doublet models (2HDMs) and composite Higgs models. Due to the apparent SM--like nature of the 125 GeV Higgs $h$, these models are forced into special limits. In particular, viable 2HDMs are 
of the ``alignment'' type: the SM limit for the $h$--couplings is achieved either (a) through decoupling of the heavy states or (b) due to special relations among the scalar potential couplings (``alignment without decoupling''). The latter case is particularly interesting since it allows for relatively light additional scalar and pseudoscalar states, although the theoretical motivation for the above coupling relations is still to be  understood. In the realm of composite Higgs models, the critical parameter controlling the deviations of the Higgs couplings from their SM values is 
\begin{equation}
\xi = {v^2 \over f^2} \;,
\end{equation} 
where $f$ is the scale of strong dynamics. To keep such deviations at the $\sim$10\% level,
$f$ is bounded from below by roughly 700 GeV, depending on the specifics of the model. New composite states are expected at the scale $g f$, where $g$ is the corresponding coupling constant.

One of the interesting applications of multi--Higgs models concerns the dark matter problem.
In 2HDMs, one may impose a $Z_2$ symmetry under which only the second Higgs doublet transforms, $H_I \rightarrow -H_I$. This forbids its coupling to fermions and makes it stable. One finds that all the dark matter constraints can be satisfied by the ``inert Higgs doublet'' $H_I$, although the direct DM detection constraints push its mass to the 500 GeV region. Sokolowska discussed an extension of this idea whereby one adds a further inert doublet. This relieves some of the constraints as long as the particle spectrum is somewhat compressed and allows for efficient co--annihilation of dark matter. As a result, new mass windows for lighter DM open up, e.g. in the regions of 400 GeV and $m_h/2$. 

Subtleties of CP transformations in multi--Higgs models were discussed by Ivanov. In general, one may accompany the usual CP transformation by a unitary transformation $U$ that mixes the Higgs doublets $\phi_i$:
\begin{equation}
\phi_i \rightarrow U_{ij} \phi_j^* \;.
\end{equation} 
This is particularly relevant when the resulting transformation is a symmetry of the potential, while the usual CP $\phi_i \rightarrow  \phi_i^*$ is not. Such a generalized CP does not necessarily square to one since $UU^*$ may not be unity. In this case, some of the (complex) scalars can be CP-half-odd instead of being odd or even, that is, they pick up the phase $i$ under the generalized CP transformation, which  leads to peculiar phenomenology.

An alternative to the common SM extensions was advocated by Shaposhnikov. The main cosmological shortcomings of the Standard Model  can be addressed by adding 3 right--handed neutrinos which would play the role of dark matter and be responsible for baryogenesis,
and including the Higgs--gravity coupling of the form
\begin{equation}
\Delta {\cal L} = -\xi_H \vert H \vert^2 R \;,
\end{equation}
where $R$ is the scalar curvature. In this case, the Higgs can play the role of the inflaton
since at large (Planckian) field values the scalar potential becomes flat allowing for slow-roll inflation. This minimalistic idea fits well with the cosmological observations and may also explain the lack of new physics signatures at the LHC and other experiments. One concern   to be aware of is some tension of the Higgs potential absolute stability with the current top--quark data, although it does not appear very significant at the moment.

\section{Flavor Physics}
 
Even though the CKM picture appears to be in excellent   agreement with the data overall (Fig.2), new physics may still affect a subset of observables. Theoretically, there is no preference for the scale of fundamental flavor dynamics: it may be at the TeV scale, in which case it must be very special, or just as well at the Planck scale as is the case in string theory. Given that the flavor structures we see in the SM $are$ special, 
 one should keep an open mind and be prepared for surprises. 
 Having said that, it is important to appreciate the sheer volume of B--physics data:
  the B$^+$ meson already has close to 500 decay modes, such that statistical fluctuations would not be unexpected.

The following topics were discussed:

\begin{itemize}
\item  B--physics anomalies
\item  progress in lattice calculations 
\item  heavy flavor production at the LHC
\item  CP violation  
\item  hadronic B--decays
\end{itemize}

\begin{figure}[t]
\begin{center}
\includegraphics[height=70mm]{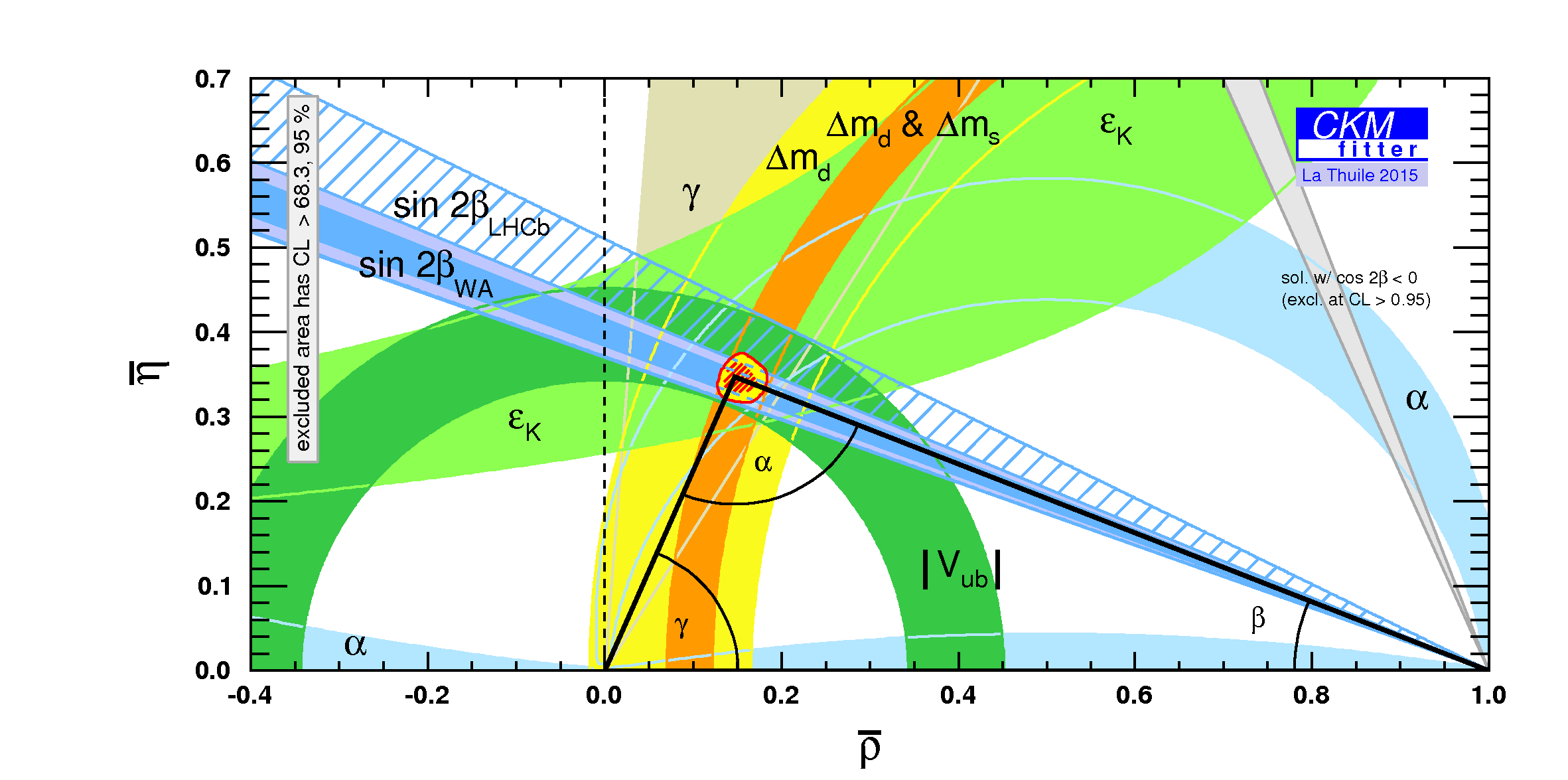}
\caption{CKM unitarity triangle. Figure credit: CKMfitter Group. }
\end{center}
\end{figure}

Flavor physics theory talks were dominated by the discussion of the current B--physics anomalies. Neubert reviewed their current status focusing on the $R_D$ and $R_K$ observables,
\begin{equation}
R_{D^{(*)}} = {   \Gamma(\bar B \rightarrow  D^{(*)}  \tau \bar \nu   ) \over     
 \Gamma(\bar B \rightarrow  D^{(*)}  l  \bar \nu   )   }~~,~~
 R_{K} = {   \Gamma(\bar B \rightarrow \bar K \mu^+ \mu^-    ) \over     
 \Gamma(\bar B \rightarrow   \bar K e^+ e^-  )   }\;,
\end{equation}
where $l= e,\mu$. These quantities are clean and sensitive probes of lepton universality in B--decays. Both of them currently exhibit deviations from the SM predictions: $R_{D^{(*)}}$
at the 3.5$\sigma $ level and $R_{K}$ at the 2.6$\sigma $ level. While the $R_{K}$ 
anomaly  is driven by the LHCb data, $R_{D^{(*)}}$ shows  deviations of varying significance in all three experiments: BaBar, Belle and LHCb. This is particularly surprising since this observable is dominated by the tree level decay in the SM and  therefore rather clean. 
In the $R_K$ case, the theoretical uncertainties are particularly small since both $\mu$ and $e$ are effectively massless, while more statistics is needed to see if the current tendency persists\footnote{The latest LHCb measurements of $R_{K^{*}}$ also show a similar size deviation from the SM prediction.}.  There are further observables that show some tension with the SM predictions,
in particular,  $P_5^\prime$ that is sensitive to the angular distributions in $B\rightarrow K^* \mu\mu$. However, the theoretical status of this  anomaly is not as firm: the hadronic uncertainties due to, for instance, breakdown of quark--hadron duality,  plague the analysis.

A possible resolution of the above anomalies in terms of new physics was offered by Virto. 
Once the operator 
\begin{equation}
{\cal O}_{9\mu} = {\alpha\over 4 \pi} (\bar s  \gamma_\mu P_L b) (\bar \mu \gamma^\mu \mu )
\end{equation}
with the Wilson coefficient $C_{9\mu}^{\rm NP}\sim -1$ is included, the quality of the fit to the data improves dramatically. Clearly, this muon--specific interaction violates lepton universality and is mostly relevant to   $R_{K}$. 
The size of the new physics contribution is quite large: in the SM the corresponding Wilson coefficient takes on the value  4.1. It is therefore  a highly non--trivial task to design a UV complete model satisfying all the flavor and as well as collider, etc.  constraints.

The case for new physics in $b \rightarrow s l^+ l^-$ transitions was strengthened by Neshatpour. He argued that introducing hadronic power corrections and fitting their
coefficients to the data including $P_5^\prime$ does not result in a better quality fit compared to that with 
just the new physics contribution $C_{9\mu}^{\rm NP}\sim -1$. As a result, this statistical view supports the new physics interpretation (modulo the above--mentioned issues).

Another  area where  possible hints of new physics effects are seen is Kaon decays. As discussed by Lunghi, advanced lattice calculations allow one to analyze the $K\rightarrow \pi\pi$ matrix elements in detail. In particular, the long standing puzzle known as the ``$\Delta I=1/2$ rule'' has been explained by a subtle accidental cancellation in the $\Delta I=3/2$ amplitude. The resulting $\epsilon^\prime/\epsilon$ exhibits some tension with the SM prediction:
  \begin{eqnarray}
&& {\rm Re} ( \epsilon^\prime/\epsilon)_{\rm exp }= (16.6\pm 2.3)\times 10^{-4} \;, \nonumber\\
&& {\rm Re} ( \epsilon^\prime/\epsilon)_{\rm th }= (1.36\pm 5.15\pm 4.59)\times 10^{-4}\;.
\end{eqnarray}
The discrepancy is at the $\sim 2 \sigma$ level, so no conclusive statement can yet be made. 

A new strategy to determine the CP violating phase $\phi_s$ of the $B_s -\bar B_s$ mixing was suggested by Vos. The Standard Model prediction for this phase is close to zero, $\phi_s \simeq -0.04$,
which makes it a convenient place to look for new physics effects. The new approach is based on the following ingredients: (a) minimal use of U--symmetry ($d \leftrightarrow s$ interchange); (b) $\gamma$ and $\phi_d$ input; (c) input of non--factorizable effects through semi--leptonic branching ratios. The resulting accuracy can reach an impressive  sub--degree level in the future.

A powerful framework to describe hadronic decays of the B--mesons was discussed by  Lu.   Within the factorization--assisted topological amplitude approach, one fits 14 non--perturbative parameters using the existing data which leads to testable predictions for more than 100 channels.  

d'Enterria discussed NNLO predictions for total bottom and charm production at the LHC. The corresponding NLO predictions have very large scale uncertainties, 30 to 60\%, which are naturally expected to be reduced at the NNLO level. Adapting the available NNLO calculations for $t \bar t$ production (Top++)  to the $b\bar b$ and $c\bar c$ final states, one finds a substantial  NNLO/NLO K--factor of about 1.2 and the anticipated reduction of the scale uncertainties. However, only the $b\bar b$ cross section appears to be  well behaved  for all PDFs  with the scale uncertainty of about 15\%. The charm production cross section suffers from large uncertainties and particular sensitivity to the low--x gluon PDF which makes the current description unsatisfactory.

\section{Precision Calculations}

Precision calculations allow us to both test the Standard Model and extract its fundamental parameters.
Among the latter, the top quark mass and the strong coupling are of particular importance. Indeed, electroweak vacuum stability depends sensitively on these parameters, especially on $m_t$ (Fig.3). One could say it is literally a ``matter of life and death''.

The topics discussed at the conference include

\begin{itemize}
 \item $\alpha_s$ extraction
\item  top mass at the LHC
\item single top production
\item PDFs
\item jet production in DIS and at the LHC
\item vector boson pair production at the LHC
\end{itemize}
Many of the presented computations were performed at an impressive NNLO level.

\begin{figure}[t]
\begin{center}
\includegraphics[height=90mm]{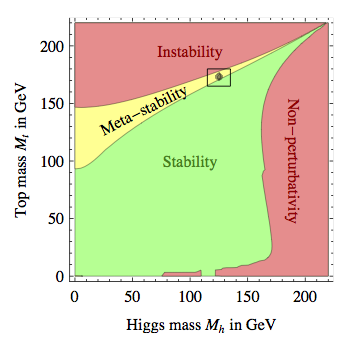}
\caption{Electroweak vacuum stability chart  (the dot indicates 
the current data).
 The figure is taken from  arXiv:1307.3536.  }
\end{center}
\end{figure}

The strong coupling affects most  LHC observables and as such is a crucial SM parameter. Its current uncertainty at $M_Z$ is just below 1\%, which leaves room for improvement.  Klijnsma reported on $\alpha_s$ determination from $t\bar t$ production in gluon fusion. This process itself at NNLO allows one to extract $\alpha_s$ with 4\% precision,
which is complementary to other measurements. Other options were  discussed by Pires (single jet production) and Niehues (DIS). In particular, Pires showed that  the NNLO calculations
for single jet inclusive production at the LHC  reduce the scale uncertainty 
to a percent level.
 Such impressive accuracy is comparable to that of the experimental measurements and will facilitate an independent $\alpha_s$ determination. Niehues   discussed NNLO	dijet	producton	in	DIS, which is also sensitive to $\alpha_s$ as well as to the gluon PDF.
There, the situation is not as transparent due to NNLO  normalization issues for the total cross section, although the scale uncertainty decreases compared to the NLO result.  
This problem can presumably be traced back to the use of the NLO PDFs.

An ingenious approach to the top quark mass measurement at the LHC was suggested by  Kim. The di--photon production in gluon fusion receives loop contributions from all charged colored particles. In particular, at the toponium threshold the di--photon spectrum exhibits a kink whose exact shape and position are determined with resummation techniques. In principle, this may allow for $m_t$ determination within 0.5 GeV.

Zhu reported on single top production at the LHC. This process provides one with a direct measurement of $V_{tb}$ among other things.  Both production and decay computations are performed at the NNLO level, with fully differential rates. Surprisingly,     the NNLO results often lie outside the NLO scale variation bands.
Further improvement is possible by going beyond the structure function approximation.

Importance of interference effects in vector boson pair production at the LHC was emphasized by R\"ontsch. At NLO, these account for a 5\% reduction of the differential rates, which is characteristic of the Higgs unitarizing behavior. The scale uncertainty at NLO decreases to the 10\% level.

Kallweit and Wiesemann discussed state--of--the--art   LHC di--boson production
at NNLO QCD. The calculations are performed  using an automated framework  ``Matrix''. The resulting NNLO corrections are found to be significant, e.g. 
the WW inclusive cross section receives a correction of 10-14\% compared to the NLO result,
while the scale uncertainty decreases to an impressive  3\% level.

One of the crucial ingredients in precision calculations is the particle distribution functions (PDFs).
Progress in PDF determination was reviewed by Nadolsky. He emphasized the importance of a coordinated effort of theorists and experimentalists to reduce the relevant uncertainties. For example, as a result of dedicated benchmarking, the uncertainty in the  gluon--gluon luminosity and thus the Higgs production was reduced from 7\% in 2012  to 3\% in 2015.  Further progress is expected from
the new generation of NNLO PDFs NNPDF3.1, CT17 and MMHT'16 which are  due to be released shortly.

Bertone discussed the xFitter project which provides a unique open--source framework to extract PDFs and  fundamental parameters  from a large variety of experimental data. For example, xFitter allows for a competitive charm quark mass determination from inclusive and charm production in DIS (HERA I+II). Another important recent application is the determination of the photon PDF from the  ATLAS 8 TeV Drell--Yan data.

Salgado reported on the update of the nuclear PDFs  EPPS16. Compared to its previous version EPS09, the new set includes the dijet and W,Z LHC data as well as neutrino results. It also supersedes  EPS09	in that the new analysis is more realistic allowing for more freedom in parametrization, which tends to  increase the PDF uncertainties.

\section{New Phenomena}

New physics at the TeV scale has traditionally been motivated by the hierarchy problem.
The quantum corrections in the SM are large and tend to drive the Higgs mass to Planckian values. However, such apparent instability may be deceiving and 
under the surface certain  stabilizing mechanisms could be at work. For instance, one may imagine that the stones in Fig.4 are kept together by a metal rod. In particle physics, the role of such a stabilizer would be played by supersymmetry (or technicolor, etc.).
On the other hand, one could also   argue that the system is simply fine--tuned  and no new mechanism is needed to explain the existence of this configuration: after all, it does not violate any laws of physics.  Given the current data, it would be premature to dismiss either of these options, even though the first possibility seems much more appealing.

The discussion in the ``New Phenomena'' session was focused around supersymmetry and composite models  (apart from the talks included in the Higgs section).

Deandrea reviewed the hierarchy problem focusing on the composite Higgs solution. If the Higgs boson is a composite state, no fundamental scalar appears at low energies and thus
the quantum corrections are well under control. This entails a tower of other composite states  \`a la $\rho$--meson to be  accessible at  (future?)  colliders. Furthermore, the Higgs interactions involve energy--dependent formfactors which can also be probed. 

  \begin{figure}[t]
\begin{center}
\includegraphics[height=90mm]{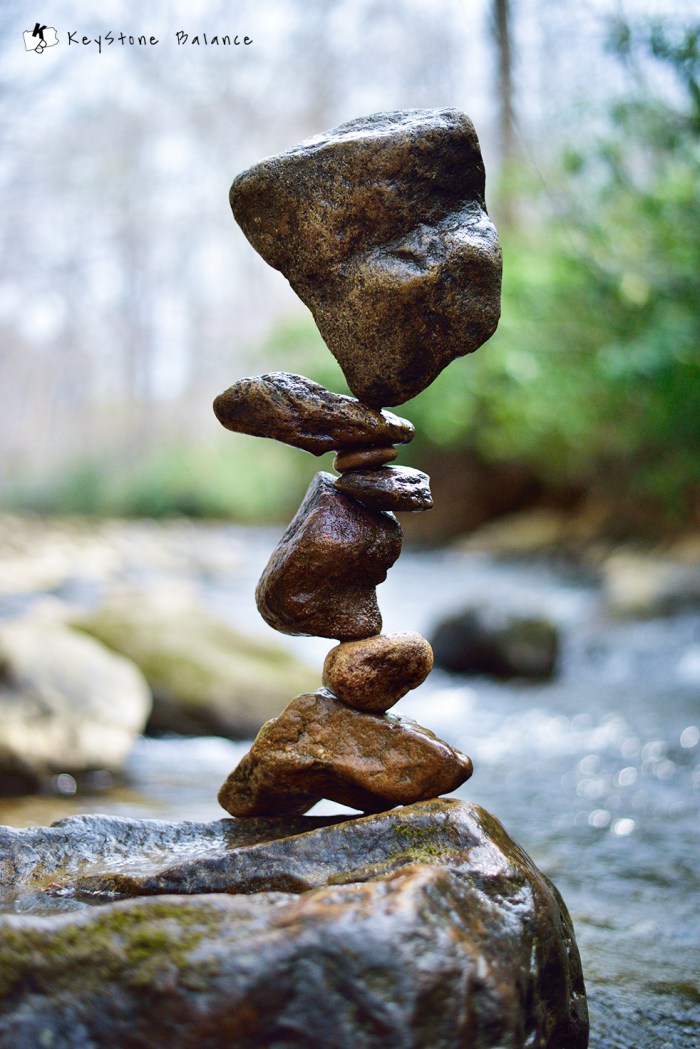}
\caption{Fine-tuned? Photo credit: Tim Anderson ({\it KeyStone Balance}).  }
\end{center}
\end{figure}

 Status of supersymmetric models was reviewed by Porod. The Higgs mass of 125 GeV requires a large radiative correction from the top quark superpartners, which implies either heavy stops or a large left--right mixing. The latter can induce charge and color breaking minima in the scalar potential, although this may not pose imminent danger. 
While typical SUSY models involve some fine--tuning to reconcile $M_Z$ with heavy superpartners, there is a 
class of general MSSMs  dubbed ``natural SUSY'' which exhibit a reduced degree of fine--tuning.
Such models  feature light SUSY particles as long as these enter the relation between $M_Z$
and the SUSY masses,
\begin{equation}
 m_Z^2 \simeq -1.8 \mu^2 + 5.9 M_3^2 + ...,
 \end{equation}
where the parameters on the right hand side are taken at the GUT scale. As a result, the higgsinos, gluinos and stops are relatively light with masses of order 100 GeV. This is consistent with the direct search constraints as long as the different higgsinos are almost degenerate (within 5--10 GeV). 

All in all, it is fair to say that  many traditional {\it pre--LHC} models are essentially excluded. 
This does not rule out more general supersymmetric and composite models: after all, when 
constructing models, we are guided by ``simplicity'' in some form while  Nature has its own ways. These frameworks address the important  issue of extreme fine--tuning to 1 part in $10^{30}$ and  should not be dismissed lightmindedly.  
Hopefully, 
 the next 15 years of the LHC   will have  more to say   about these theories.

 \section{Summary}

  The highlights and take--home lessons    from  Moriond QCD 2017
 can be summarized by the following bullet points:
 \begin{itemize}
 \item {\bf remarkable progress in precision calculations.} NNLO precision is  now a commonplace allowing for unprecedented accuracy  at a hadron machine. 
\item  {\bf Higgs precision era.} Higgs $p_T$ distributions, interference effects, etc. 
allow for accurate tests of the Higgs nature.
\item  {\bf intriguing B--physics anomalies.} Several $2-3\sigma$ deviations in clean  semi-leptonic observables leave ample room for new physics.
\item  {\bf data--driven theory: test ``unmotivated'' ideas.} Given the absence of striking signatures of ``traditional'' forms of new physics, the theory approach should be more inclusive. May require painstaking analysis of difficult observables.  
\end{itemize}

{\bf Acknowledgements.} I am very grateful to the organizers, especially Nazila and Maria, for their kind invitation
which allowed me to spend some (limited, for obvious reasons) time on the beautiful slopes of La Thuile.

\end{document}